\documentstyle[twocolumn,floats,aps,times]{revtex}

\input epsf
\epsfverbosetrue

\draft

\begin{document}

\draft

\title{Interference and Interaction in Multiwall Carbon Nanotubes}
\author{C.~Sch\"onenberger$^{\dagger}$, A.~Bachtold$^{\dagger}$,
C.~Strunk$^{\dagger}$, J.-P. Salvetat$^{\ddagger}$, and
L. Forr\'o$^{\ddagger}$}
\address{$\dagger$ Institut f\"{u}r Physik, Universit\"{a}t Basel, 
Klingelbergstr. 82, CH-4056 Basel, Switzerland \\
$\ddagger$ Institut de G\'enie Atomique, \'Ecole Polytechnique
F\'ed\'erale de Lausanne, CH-1015 Lausanne, Switzerland}
\date{\today}
\maketitle

\begin{abstract}
We report equilibrium electric resistance $R$
and tunneling spectroscopy 
\mbox{($dI/dV$)} measurements 
obtained on single multiwall nanotubes
contacted by four metallic Au fingers from above.
At low temperature quantum interference
phenomena dominate
the magnetoresistance. The phase-coherence
\mbox{($l_{\phi}$)} and elastic-scattering lengths
\mbox{($l_e$)} are deduced. Because $l_e$
is of order of the circumference of the nanotubes, 
transport is quasi-ballistic. This result is
supported by a \mbox{$dI/dV$} spectrum which
is in good agreement with the density-of-states (DOS) 
due to
the one-dimensional subbands expected
for a perfect single-wall tube.
As a function of temperature $T$ the
resistance increases on decreasing $T$
and saturates at \mbox{$\approx 1-10$\,K} for
all measured nanotubes. 
$R(T)$ cannot be related to the energy-dependent
DOS of graphene but is mainly
caused by interaction and 
interference effects.
On a relatively small voltage scale of order 
\mbox{$\approx 10$\,meV}, a pseudogap
is observed in $dI/dV$ which agrees with
Luttinger-Liquid theories for nanotubes.
Because we have used
quantum diffusion based on 
Fermi-Liquid 
as well as Luttinger-Liquid theory
in trying to understand our results, a large
fraction of this paper is devoted to a careful
discussion of all our results.
\end{abstract}

\vspace{.2cm}
\vspace{.2cm}

\noindent
{\large\bf 1 Introduction}\\

In recent years carbon nanotubes have suprised us
with remarkable properties  
which are neither present in graphite
nor in diamond\cite{ReviewNT}.
Nanotubes are ideal
model systems 
for the investigation of
low-dimensional molecular conductors
whose electronic properties are 
largely determined by molecular orbitals 
that are extended along the whole conductor
similar to metals.
The length-to-diameter ratio
can exceed $10^5$ for the
smallest diameter tubes
(\mbox{$\agt 100$\,$\mu$m} in length and 
\mbox{$\approx 1$\,nm} in diameter).
A carbon nanotube is obtained from a slice
of graphene
sheet wrapped into a seamless cylinder. Depending
on the specific realization (the wrapping vector), the nanotube
may be a true one-dimensional metal with a non-vanishing
density-of-states at the Fermi energy or a semiconductor
with a gap.
This is in marked contrast to the two-dimensional graphene
sheet which is a zero-gap semiconductor.
By combining metallic and semiconducting tubes, either in
their intrinsic or doped forms, the whole span of 
electronic components ranging from wires, bipolar devices
to field-effect transistors may be embodied in nanotubes.
A striking field-effect has already been demonstrated
at room temperature\cite{NTFET}. On the fundamental side, a perfect
metallic nanotube is supposed to be a ballistic conductor
in which only two one-dimensional ($1d$) subbands (modes)
carry the electric current\cite{NT2Modes}. 
Because of the relatively low
carrier concentration as compared to ordinary metals like
Au, strong correlations due to Coulomb interactions are
expected\cite{NTEEinteraction,NTLL1}.

Nanotubes can be fabricated in two basic forms: either
as single-wall cylinders (SWNT = single-wall nanotube)
or as wires that consist of a set of concentrically arranged
cylinders (MWNT = multi wall nanotube).
Furthermore, both forms may be present as single wires
or packed into ropes.
A remarkable variety of physical phenomena have been observed
in electrical transport to date. 
The first signature of quantum effects were found in
the magnetoresistance (MR) of MWNTs.
Song~{\it et al.} studied 
bundles of MWNTs\cite{Song94} 
while Langer~{\it et al.} was able to measure
the MR of a single MWNT for the first time\cite{Langer}.
In both cases a negative MR was observed at 
low temperatures
indicative of weak localization. 
However, the phase-coherence length $l_{\phi}$
was found to be small 
amounting to only \mbox{$\alt 20$\,nm}
at \mbox{$0.3$\,K}, in strong contrast
with the 
ballistic transport with
only $2$ modes 
theoretically expected
for a perfect nanotube. 
Evidence for much larger
coherence lengths in SWNTs was provided by 
the observation of zero-dimensional states in 
single-electron
tunneling experiments\cite{Tans,NTSET}.
Recently,
a pronounced Aharonov-Bohm resistance
oscillation has been 
observed in MWNTs\cite{BachtoldAB}. This
experiment has provided
compelling evidence that $l_{\phi}$ can exceed 
the circumference of the tube so that large coherence
lengths are possible for MWNTs too, see also\cite{BachtoldKirchberg}.
Because the
magnetic-flux modulated resistance agreed
with an Aharonov-Bohm flux
of $h/2e$ the effect is supposed to be caused 
by conventional weak-localization for which
backscattering of electrons is essential. 
In essence, as in the work of Langer~{\it et al.}\cite{Langer},
$2d$-diffusive transport could explain 
the main observation
reasonably well. The
Aharonov-Bohm experiment lead to a convincing
proof that the electric current 
flows in the outermost (metallic) graphene 
tube only, at least at
low temperatures \mbox{$T \alt 70$\,K}. 
Presumably, this
is a consequence of the 
way in which the nanotubes are contacted.
Electrodes 
are evaporated over the MWNT 
and therefore contact 
the outermost tube preferentially.
Because it is only the
outermost tube that carries 
the current, large diameter single
graphene cylinders can now be investigated.
Very recently, proximity induced
superconductivity was found in weak-links formed
by a nanotube in contact with two superconductig
banks\cite{NTweaklink}.

All these striking results mentioned
were obtained by
contacting a single nanotube with the aid of
micro- and nanostructuring technologies.
Alternative approaches have been developed
as well. For example, 
Dai~{\it et al.} and Thess~{\it et al.} 
have measured the voltage drop
along nanotubes
using movable tips\cite{Dai96Thess96} and
Kasumov~{\it et al.}
have developed a pulsed-laser deposition
method\cite{Kasumov98}. 
Furthermore, scanning-probe manipulation schemes
were developed\cite{Pablo99Shea99} and recently is
has been shown that SWNTs can directly be
synthesized to bridge pre-patterend 
structures\cite{Kong98}.
Still another elegant 
method allowing to
electrically contact a single MWNT
has been used by Frank~{\it et al.}\cite{Frank}.
The nanotube,
which is attached to a tip, is contacted by immersing
it into a liquid metal like mercury. Immersing and
pulling out the nanotube repeatedly is claimed
to have a cleaning effect.
In particular, graphitic particles
are removed from the tubes. After some
repetitions, an almost universal conductance
close to the quantized value $G_0:= 2e^2/h$
is measured. From these experiments the
researchers conclude that transport in MWNTs is
ballistic over distances of the order of
\mbox{$\agt 1$\,$\mu$m}. This is very striking
because the experiments were conducted at
room temperature. At present it is not clear
why the conductance is close to $G_0$
instead of the theoretically expected value
of $2G_0$.

Our results, which were 
obtained on similarly arc-produced MWNTs,
appear to be in strong 
contradiction to the ballistic transport of 
Frank~{\it et al.}\cite{Frank}. 
It is one aim of this
paper to address this contradiction by quantifying
the degree of backscattering (or diffusivness)
observed in our samples due to static disorder. 
As will be demonstrated,
the discrepancy is much weaker than originally thought.
It turns out that our nanotubes may be regarded
quasi-ballistic conductors, at least
at low temperature. At room temperature
the degree and source of backscattering is difficult
to quantify. But because the measured resistances
are in quite good agreement 
with theoretically predicted
values backscattering must be relatively weak. 

The fabrication of reliable electric contacts
to single carbon nanotubes 
has been
an issue of great attention during recent years
and quite some progress 
has been 
made\cite{Tans,NTSET,BachtoldAB,BachtoldKirchberg,NTweaklink,Dai96Thess96,Kasumov98,Pablo99Shea99,Kong98,Frank,BachtoldAPL,Ebbessen,Muster98}.
The same
remark is appropriate for the quality of the
nanotubes. The electric contact resistance
of SWNTs and MWNTs, which were adsorbed on
a prestructured electrode pattern, were
found to be large, in the \mbox{M$\Omega$} 
range\cite{Tans,BachtoldKirchberg,BachtoldAPL}.
The contact resistance can be reduced
by local electron-beam irradition enabling to
select a single nanotube
for measurement\cite{BachtoldAPL}.
If the contacts are high-ohmic, they
determine the measured two-terminal 
resistance.
Under such circumstances
four-terminal measurement
schemes are usually applied in order to
measure the intrinsic resistance.
The first four-terminal resistance measurements
on single MWNTs were realized by 
Ebbessen {\it et al.}\cite{Ebbessen}.
All kind of temperature
dependences were found. Since resistivities
at \mbox{$300$\,K} varied by six orders of magnitude,
from \mbox{$5\cdot 10^{-6}$} to \mbox{$6$\,$\Omega$cm},
this was taken as evidence that nanotubes can
be metallic or semiconducting. However, these results 
suffer from possible damage caused by the
method used to structure the devices
(ion-induced deposition using a metalorganic precursor gas).
Much lower contact resistances can be obtained
if the contacts are evaporated directly onto the
nanotubes.
Our own record is presently at
\mbox{$R_c \alt 100$\,$\Omega$}.
Note, that we define the contact resistance by
$R_c:=(R_{2t}-R_{4t})/2$, where $R_{4t}$ denotes
the four-terminal resistance (outer contacts are
used for current source and drain, whereas the
two inner ones serve to measure the voltage drop) and
$R_{2t}$ denotes the two-terminal resistance measured 
using only the two inner contacts.

At this point an important remark has to be made.
If we assume that the nanotubes are indeed nearly
perfect, for the sake of the argument, say
ballistic, a quantized resistance is expected
providing the contacts are ideal (no backscattering
off the contacts). For the nanotube with two
propagating modes the resistance would then equal
$h/4e^2$. This resistance is known to be a pure
contact resistance which originates from 
distributing the current carried by a few modes 
into many modes of a  macroscopic
reservoir\cite{QuantumContactR}. 
Hence, once the
original goal to lower apparent contact resistances
(i.e. $R_c$) has been achieved,
we end up measuring in either
two- or four-terminal configuration 
the same Landauer-B\"uttiker\cite{Landauer,Buettiker}
quantized 
contact resistance.
Because the distinction between
two- and four-terminal resistance disappears
for ideal contacts, one
could equally well step back to the original two-terminal
configuration in this situation.
In reality contacts are not expected
to be perfect: most importantly, 
the contact is abrupt and 
therefore everything else 
then an adiabatic
widening which is realized in semiconductor 
split-gate point contacts\cite{QPC}. 
Hence, we expect that each contact
attached to a nanotube will inevitably induce some
degree of backscattering. 
In such a realistic case, it is 
difficult to relate the measured resistance with
either the `intrinsic' resistance or the one
that would be measured if the contacts were ideal.
For the experiments we are going to discuss, 
$R_c$ is in the range of $1$ to 
\mbox{$10$\,k$\Omega$}.
Because this is comparable
with the quantized resistance 
of an ideal nanotube,
one should keep in mind that
absolute resistance values may be off by a 
factor $\approx 2$. In the next
section we will come back
to this point and report recent
room temperature 
resistance values.
 
Because we are reporting electric transport
measurements for MWNTs, we would like to
bring the differences and similarities 
between MWNTs and SWNTs to
the reader's attention.
Since the electrical current of contacted MWNTs
has been demonstrated to flow through the outermost
carbon tube at low temperatures\cite{BachtoldAB},
experiments using
MWNTs also address a single shell. The main difference
is the diameter of the respective graphene cylinders.
SWNTs have a typical diameter of
\mbox{$\approx 1-1.5$\,nm} while our
MWNTs have diameters in the range \mbox{$10-20$\,nm}.
This one order of magnitude difference in
diameter relates into an order of magnitude
difference in energy scale.
For a semiconducting nanotube,
tight-binding calculations predict a gap
of order \mbox{$0.1$\,eV} for a \mbox{$10$\,nm}
diameter tube, whereas for a SWNT the gap
is of order \mbox{$1$\,eV}\cite{AijkiAndo}.
This energy
scale is also roughly valid for 
the energy separation
between the first conduction and valence
subbands above and below the
Fermi level for metallic tubes.
Differences in electric
resistance due to either semiconducting
or metallic behavior 
are expected to be more pronounced
for SWNTs because of
the larger energy scale. In fact, these
differences were observed with tunneling 
spectroscopy on SWNTs at low temperature\cite{Wildoer}.

If a SWNT is adsorbed on a substrate with
metallic contacts, the nanotube has been
found to closely follow the contour of the
surface\cite{followcontour}.
Because the metallic contacts
have a finite thickness, the SWNT is strongly
bent just at the edge of the contacts.
These bending defects have been suggested to
be one reason for the very high-ohmic contacts
found in SWNTs contacted in this way\cite{followcontour}.
This also suggests that
all kinds of local deformations like bends, kinks,
and twist will cause some elastic backscattering,
hence will destroy the exact quantization of
the resistance valid for an ideal nanotube
with transmission probability $t=1$.
Because all surfaces exhibt in general some
roughness on the nm-scale (except for atomically
flat well prepared substrates) it is hard to
imagine that electric transport is ballistic
for SWNTs adsorbed on a substrate.
Here MWNTs (and also ropes of SWNTs) are advantageous,
because the outer electrically measured tube
is supported by $10-20$ additional tubes inside
which enhance the tube's ridgidity. Computer simulations
show that SWNTs may even collapse due to
adhesive forces,
whereas MWNTs are hardly deformed\cite{NTcollapse}.
Furthermore,
highly transparent contacts are presuambly
easier to achieve in case of MWNTs because
the larger diameter results in a larger contact
area for electric contact fingers of a 
given width\cite{ContactSimulation}.

There are other reasons which are in favour of
MWNTs. Large diameter carbon nanotubes have been
predicted to have a larger mean-free path $l_e$
for elastic scattering for a given 
defect density assuming a two-band 
tight-binding
model with fluctuating diagonal and 
off-diagonal matrix elements\cite{WhiteNature}.
The variance $\sigma$ of these matrix elements describes 
the strength of the disorder.
The mean-free path is found to be proportional to
the diameter $d$ 
as long as the subband separation is larger than
the energy broadening due to scattering. This
relates into the condition
that $\sigma < C/\sqrt{d}$ where $C$ is a constant.

As mentioned before, 
Frank~{\it et al.} have suggested that transport
in MWNTs is ballistic over micrometer distances
even at room temperature.
But in our opinion, no direct measurement of
the elastic-scattering length exists for either
MWNTs and SWNTs.
Due to the high contact resistances, transport in
SWNTs is largely dominated by single-electron
tunneling effects\cite{Tans,NTSET}.
Here, charge transport is dominated by the
contacts and the Coulomb energy related
to adding (removing) electrons to (from) 
the nanotube as a whole. Transport within the
nanotube is of little significance for the
externally measured resistance. Spectroscopy allows
to access discrete $0d$-states whose width can be
used as a measure for the dephasing rate
(but not for the rate of elastic scattering).
Using this method, SWNTs have been demonstrated
to be phase-coherent conductors over large
distances \mbox{$> 1$\,$\mu$m} at low
temperature \mbox{$\alt 1$\,K}\cite{Tans}.

SWNTs have an additional disadvantage: their 
large curvature 
turns all nanotubes into 
semiconducting ones
except for armchair nanotubes.
This curvature effect can safely be neglected for
MWNTs. In contrast, for MWNTs one may have
to take a weak hybridization of electron
states from neighboring
tubes into account. Because of the
different diameters, the respective
graphene lattices can never
match up commensuratly.
Hybridization is therefore expected to
be small. Note however, that in transport
of MWNTs the
energy landscape of an neighbouring 
nanotube acts as a random potential. Taking
an intertube overlap energy of 
\mbox{$0.25$\,eV}
we estimate (based on the work of 
White and Todrov\cite{WhiteNature})
that \mbox{$l_e$} can still exceed \mbox{$1$\,$\mu$m}
for nanotubes with a diameter of 
\mbox{$30$\,nm}.

With regard to electrical measurements in magnetic
fields, orbital effects cannot be studied in SWNTs
because extremely high magnetic fields would be required.
In contrast, a magnetic field of \mbox{$12$\,T}
is enough to induce a magnetic flux of $h/2e$
in a MWNT with outer diameter \mbox{$15$\,nm}.
For this reason, valuable information
can be obtained from magnetotransport
experiments of MWNTs in parallel and perpendicular
field. The Zeeman energy splitting
due to the magnetic moment of the electrons
can of course be observed both in single and
multiwall nanotubes. 
From measurements
of the Zeeman energy of SWNTs the gyromagnetic
ratio $g$ was found to be very close to the
free electron value, i.e. $g=2$\cite{Tans,NTgfactor}.
Finally, the larger size of MWNTs (and SWNT ropes)
facilitates
their handling\cite{Avouris} 
and imaging by conventional
scanning electron microscopy (SEM).

This brief advertisement for MWNTs 
should make clear that the structure and 
large diameter
of MWNTs can be advantageous for the
studying of certain electrical properties of carbon
nanotubes. From a theoretical point of view,
single-wall nanotubes are the model systems
of choice. However, since MWNTs do behave
as single tubes at low temperature, theories are equally
well applicable to MWNTs.

A perfect metallic nanotube (also the large
diameter ones) has two $1d$-subbands at the 
Fermi energy $E_F$ originating from the
$\pi$ and $\pi^{\star}$ bands of graphene.
The graphene bands intersect exactly
at the corner points of the  Brillouin zone, 
known as the $K$ points.
In the vicinity of these points the energy
of the bands are given by
$E(\vec{k})= \pm\hbar v_{F}|\vec{k}-\vec{K}|$,
where the Fermi velocity $v_F$ is 
\mbox{$\approx 10^6$\,m/s}\cite{GrapheneReview}.
Due to the two available subbands
at $E_F$, nanotubes should ideally have
a conductance of 
\mbox{$G=4e^2/h= (6.4$\,k$\Omega)^{-1}$}.
No clear evidence for this universal
(independent of size and wrapping)
conductance for metallic tubes have been
found to date. Frank et al. find values
close to $2e^2/h$ and sometimes even some
pre-plateaus at $e^2/h$\cite{Frank}.
There are also
no systematic measurements on single
nanotubes from which
both the elastic $l_e$ and the dephasing
length $l_{\phi}$ can be extracted with the
exception of the the pioneering work by
Langer {\it et al.}\cite{Langer}.

This work intends to fill this gap. 
Values for
$l_e$ and $l_{\phi}$ have been deduced
from measured interference effects
such as weak localization (WL) and 
universal conductance fluctuations (UCF).
We use conventional theories developed
for diffusive transport appropriate
for wires containing
many conducting channels and discuss the
shortcomings of this approach at the end.

Long-range Coulomb interaction
of electrons in a one-dimensional wire
(few conducting modes only) should 
strongly modify the
Fermi liquid picture of quasi-particles.
An appropriate effective description
is believed to be given 
by the Luttinger liquid (LL) model\cite{NTLL1,NTLL2,NTLL3}.
There is now striking evidence 
for LL-like behavior
in SWNTs\cite{Bockrath}. 
What about MWNTs?
Can electron-electron interaction
be accounted for by conventional
Fermi liquid theory
or do we have to use 
the LL-picture for MWNTs too?

\vspace{5mm}
\noindent
{\large\bf 2~Experimental, Results and Discussion}\\

\noindent
{\large\em 2.1~Contacting single nanotubes}\\

Single multiwall carbon nanotubes (MWNTs) are contacted
using conventional nanofabrication technology.
A droplet of a dispersion of 
arc-produced nanotubes (NTs)
in chloroform 
is used
to spread the NTs onto a piece of 
thermally oxidized \mbox{($400$\,nm)} Si(100)
substrate. Then, a PMMA
resist layer is spun over the sample. An array
of electrodes, each consisting of four contact fingers
together with bonding pads, is exposed by electron-beam
lithography. After development,
a \mbox{$120$\,nm} Au film is thermally evaporated in
a high-vacuum chamber.
After lift-off, each structure of the array
consists of four thin and
narrow Au fingers (\mbox{$\approx 100-200$\,nm} in width)
that end in bonding pads.
We emphasize that no adhesion layer like Cr or Ti 
has been used.
This ensures that no magnetic impurities
are introduced. The sample is now inspected
by SEM and the structures that have one single
nanotube lying below all four electrode fingers
are selected for electric measurements.
An example of a single MWNT contacted by four
Au fingers with separation
\mbox{$350$\,nm} (center-to-center)
is shown in Fig.~1.
Since the success of this contacting 
scheme works by chance, it is obvious that the 
yield is low. There are many structures which
have either no or several NTs contacted 
in parallel. However, since a large array of
more than $100$ structures can readily be
fabricated, this scheme has turned out to be
very convenient.
Alternatively, it is also possible to first
structure a regular pattern of 
alignment marks on the
substrate. After adsorbing the NTs, the
sample is first imaged with SEM in order to
locate suitable NTs. Having notified the coordinates
aided by the marks, the electrodes
can be structured directly onto the respective
NTs with high precision. This improves
the yield at the cost of an additional 
lithography step.
\begin{figure}[htb]
  \epsfxsize=75mm
  \centerline{\epsfbox{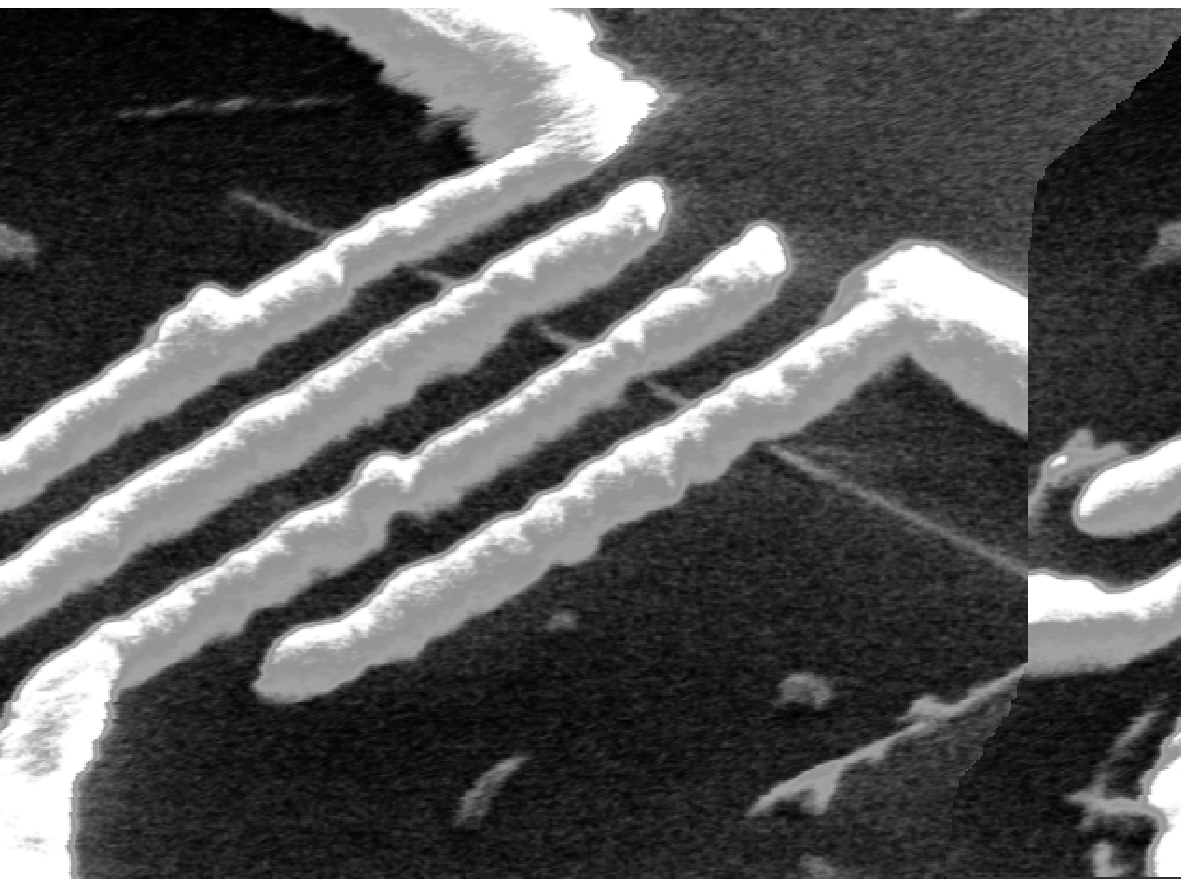}}
  \vspace{9pt}
  \caption{
  Scanning-electron microscopy image of a single multiwall
  nanotube (MWNT) electrically contacted by four Au fingers from
  above. The separation between the contacts is
  \mbox{$350$\,nm} center-to-center.}
\end{figure}  

Let us emphasize that the Au electrodes
are evaporated directly onto the nanotubes.
Previously, we have been using the reverse
scheme in which the electrode structures are made
first and the NTs are adsorbed thereafter.
In this latter scheme (nanotube over the contacts)
the contact resistances were found to be large
\mbox{($\agt 1$\,M$\Omega$)}.
It was only with the aid of local electron
exposure directly onto the NT-Au contacts that
this resistance could be lowered
to acceptable values\cite{BachtoldAPL}.
In contrast, surprisingly low 
contact resistances
(\mbox{$\approx 5$\,k$\Omega$})
are obtained with the former scheme (nanotube under
the contacts). It is also important to note that 
our Au-NT contacts
are not long-term stable at room temperature. The contacts
between Au fingers and NT are lost after a period
of typically $1-2$ weeks.
Hence, it
is important to measure the samples immediately
after fabrication.

Only samples in which a single NT 
is contacted are selected for 
conventional equilibrium resistance measurements.
An additional
selection criterion based on contact resistances
has been applied too:
we require that all contact resistances are  
\mbox{$\alt 10$\,k$\Omega$} at room temperature. 
All NTs satisfying these selection
criteria appear to be metallic, because
the resistance is found to saturate in the
limit of low temperatures.
It is quite plausible that a semiconducting tube
may never yield low-ohmic contacts due to the
formation of a Schottky barrier.
If this hypothesis is true we possibly throw away
the semiconducting tubes and retain only 
the metallic ones for measurement.
However, we are puzzled by the observation 
that of all contacted single 
NTs $80\%$ have
at least two low-ohmic contacts over
which a two-terminal 
resistance \mbox{$\alt 10$\,k$\Omega$}
is measured. 
Hence, one could
conjecture that $80\%$ of our MWNTs 
are apparently metallic. This 
contradicts the expectation
that $2/3$ of all NTs are semiconductors
provided all diameters and chiralities
are present with equal probability.

From a large number of MWNTs, 
each with four
low-ohmic contacts, the average
contact resistance is \mbox{$3.8$\,k$\Omega$} with
a standard deviation of \mbox{$5$\,k$\Omega$}
at room temperature ($R_c$ ranges from
\mbox{$100$\,$\Omega$} to \mbox{$20$\,k$\Omega$}).
With our increase of experience in fabricating
contacts, $R_c$ could be lowered gradually.
Our most recent batch shows values in the range
of \mbox{$100$\,$\Omega$} to \mbox{$1$\,k$\Omega$}.
In this limit of very low-ohmic contacts
$R_{2t}\sim R_{4t}$, as expected for
nearly ideal contacts.
An ideal
contact is defined to have no backscattering
and to inject electrons in all modes equally.
Electrons incident from the NT to the contact
will then be adsorbed by the contact 
with unit probability.
Because the contact couples to both right
and left propagating modes equally, Ohm's law
should be valid in this limit. It is important
to realize that for ideal contacts
$R_{4t}$ cannot contain any
nonlocal contribution, i.e. contribution to
the resistance that arise from a NT segement not
located in between the inner two contacts.
Any sign of nonlocality points to the presence
of nonideal contacts.

Let us emphasize again the distinction between
$R_c:=(R_{2t}-R_{4t})/2$ and the quantized 
conductance of an
ideal NT. Theoretically, the conductance of an ideal NT
is quantized: \mbox{$G=4e^2/h$}.
The origin of this resistance can be traced back
to the interface between 
the two-mode wire and the 
ideal macroscopic contact. In this
respect it is purely a contact resistance\cite{QuantumContactR}.
This quantized resistance should not be confused with
$R_c$. In the limit of an ideal NT with ideal
contacts $R_c=0$, but $R_{2t}=R_{4t}=h/4e^2$.
 
There are also samples which by chance have just one
relatively high-ohmic contact, i.e.
\mbox{$R_c=0.1-1$\,M$\Omega$}. Such samples
are used for tunneling spectroscopy. The voltage
dependent differential conductance is measured
on this contact. Because this contact
resistance is $1$ to $2$-orders
of magnitude higher than the
resistance of the NT, 
most of the applied voltage drops 
locally at the contact. Such measurements 
give information
on the local electron density-of-states and are
complementary to the transport experiments.
It would be highly desirable to develop a scheme
that allows the controlled fabrication of both low
and high-ohmic contacts to the same NT.

Finally, we mention that 
a single high-ohmic contact has 
occasionally been used
as a gate electrode. However,
the resistance of our metallic nanotubes
could barely be modified by the gate voltage,
indicating that Coulomb blockade is 
absent. This is presumably due to the
other low-ohmic contacts
that couple the
NT strongly to the environment.

\vspace{5mm}
\noindent
{\large\em 2.2~Temperature dependent resistance}\\

Four-terminal electrical resistances $R$ have 
been measured
as a function of temperature $T$ in a He-3 system
down to \mbox{$T\approx 0.3$\,K}. The resistance
always increases with decreasing temperature
and saturates around \mbox{$1-10$\,K}. 
A typical
example is shown in Fig.~2
(the dashed line corresponds to 
\mbox{$h/2e^2=12.9$\,k$\Omega$}).
The increase from room temperature is moderate
amounting to a factor $\alt 2-3$. This together
with the 
low-temperature saturation 
is taken as evidence for
the metallic nature of the MWNTs.
Non-metallic behavior is characterized
by a diverging resistance for \mbox{$T\rightarrow 0$}
as observed, for example, for semiconductors
and in more exotic transport regimes
like variable-range hopping
and strong localization.
We emphasize that not only is the temperature dependence
of $R(T)$ similar for all samples,
but the absolut resistance values also
fall into a relatively narrow range of
\mbox{$R_{4t}\approx 2-20$\,k$\Omega$}, quite different
to the original results of Ebbessen 
{\it et al.}\cite{Ebbessen}.
\begin{figure}[htb]
  \epsfxsize=75mm
  \centerline{\epsfbox{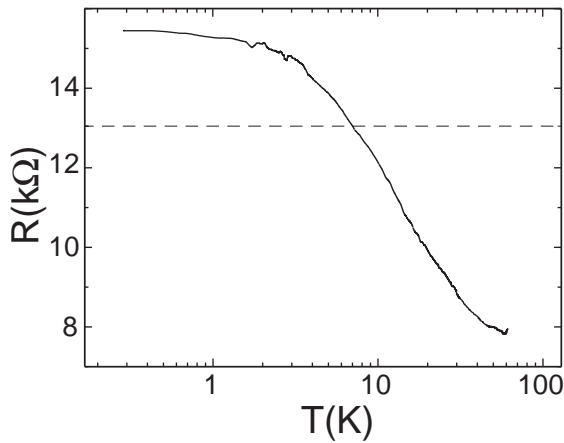}}
  \vspace{9pt}
  \caption{
  Typical temperature dependent electrical resistance $R(T)$ of
  a single MWNT measured in a four-probe configuration, i.e.
  the current is passed through the outer contacts and voltage is
  measured over the inner ones. The dashed line corresponds
  to the resistance quantum $h/2e^2$.}
\end{figure}  

The resistance increase at low temperatures
is markedly different to what is known from
(HOPG) graphite\cite{DresselhausSpringer}.
There, the resistivity decreases 
with decreasing temperature as commonly associated
with metallic behavior. The decrease is caused
by the suppression of electron-phonon scattering
at low temperature. Although nanotubes are
composed of the same graphene sheets,
the $R(T)$-dependences are different. Where does this
difference come from?

In trying to understand the temperature
dependence of $R$, we first consider
the simplest possible model.
We will compare the absolute measured resistance 
values with an expression for the classical
Drude resistance taking one graphene cylinder
and assuming $2d$-diffusive transport, i.e.
assuming $l_e \ll \pi d$.
We thereby completely disregard
the quantization of the wavevector
around the tube circumference leading to
$1d$ subbands.
The electron states have energy
$E=\pm \hbar v_F|\vec{k}|$, where $\vec{k}$
is the $2d$-wavevector measured with respect
to the two independent Brillouin corner
points. The Fermi energy is taken to be
\mbox{$E_F=0$} and a reasonable value
for the Fermi velocity 
\mbox{$v_F=10^{6}$\,m/s} is assumed\cite{Tans,GrapheneReview}.
For the electron density-of-states (DOS) we obtain
$n_{2d}(E) = 2E/\pi(\hbar v_F)^2$. Using the
Einstein equation
$\sigma_{2d} = e^2 n_{2d} D$,
which relates the conductivity $\sigma_{2d}$
to the diffusion coefficient $D=v_F l_e/2$
and the electron density, 
the energy-dependent conductivity
is found to be $\propto E$.
To obtain the equilibrium sheet conductivity
$\sigma _{2d}$
at finite temperature 
the energy has to be replaced by $kT$. 
The equation now reads:
\begin{equation}
  \sigma_{2d} \approx \left(\frac{2e^2}{h} \right)\frac{kT}{\hbar v_F}l_e
\end{equation}
Due to the vanishing electron DOS
for $E\rightarrow 0$,
the resistance of 
a graphene sheet increases with decreasing temperature
following $R(T)\propto T^{-1}$.
Although a resistance increase is observed,
the increase is not compatible with a $T^{-1}$
dependence.
Moreover, $R(T)$ saturates below \mbox{$\approx 4$\,K}.
This saturation could be explained by a finite
overlap with additional graphene cylinders
giving rise to a narrow band 
of width $\Delta$ at the Fermi energy
as in graphite.
In the limit $kT < \Delta$ a constant
DOS would develop.
Now, $kT$ has to be replaced by $\Delta$
in the above equation for $\sigma_{2d}$.
Let us put in numbers in order to estimate the
mean-free path $l_e$. Taking from the 
experiments 
\mbox{$R=10$\,k$\Omega$}, contact
separation \mbox{$L=350$\,nm}, tube diameter
\mbox{$d=20$\,nm} and \mbox{$\Delta=4$\,K}
(as suggested by the $R(T)$ saturation),
we obtain \mbox{$l_e\approx 13$\,$\mu$m}.
This large mean-free path violates the
assumption that diffusion is $2$-dimensional.
Even more serious, $l_e \gg L$. The only
way to reconcile this model with the
requirement $l_e \alt L$ is to assume that
a large number ($30$) of graphene cylinders
carry the electric current equally. 
We know from the Aharonov-Bohm experiments
that this is not the case\cite{BachtoldAB}.
We therefore conclude that the specific
temperature dependence of $R$ cannot be
related to the energy-dependent
DOS of graphene.

Within this simple Drude picture, 
the discrepancy can be
resolved if we take into account
the band-structure modifications imposed
by the periodic-boundary condition along
the circumference of the cylinder leading
to $1d$-subbands. In contrast to graphene,
for which the DOS tends to zero
for $E\rightarrow 0$, the DOS
is constant in an relatively large energy window
centered around the Fermi energy.
This energy window is given by the
subband separation $\Delta E_{sb}$.
Instead of the hypothetical 
and small hybridization 
energy,
$\Delta E_{sb}$ should be inserted in the previous
equation. With
\mbox{$\Delta E_{sb}=100$\,meV}, typically valid
for the outermost cylinders of our MWNTs,
one arrives at a mean-free path
of \mbox{$l_e\approx 50$\,nm}, which is of
order of the circumference of the tube.
This number is of reasonable
magnitude and in agreement with
magnetoresistance measurement
(see Sect.~2.4). This argument suggests that
electron transport in MWNTs
is not $2d$-diffusive, but rather 
one-dimensional. Most importantly,
it demonstrates that the $1d$-subbands
need to be considered in MWNTs as well.

The classical $1d$-Drude resistance due to
static-disorder alone 
predicts a temperature-independent
resistance. Temperature dependences can be
caused by other scattering mechanism,
like electron-phonon and electron-electron 
scattering.
In a first approximation we assume 
that electron-phonon
scattering does not contribute significantly
to momentum relaxation in carbon nanotubes.
Since otherwise, R(T) should decrease with decreasing
temperature as observed for graphite.
It is known however that 
electron-electron interaction 
can contribute significantly to $R$ in nanoscaled devices
at low temperature.
In the limit of a piece of metallic wire
which is weakly coupled to the environment, i.e.
with \mbox{$R_c\gg R_Q$}
\mbox{($R_Q=h/2e^2=12.9$\,k$\Omega$} 
is the resistance quantum), Coulomb blockade 
turns the system into an insulating
state\cite{CBreview}.
This is the case, if $kT \ll E_c$, where
$E_c=e^2/2C$ is the single-electron charging 
energy of the island (wire). The capacitance
$C^{\prime}$ per unit length
for our MWNTs is estimated from the
expression of the geometric capacitance
valid for an infinite conducting
cylinder with diameter $d$ supported above
a conducting backplane at distance $a \gg d$:
\begin{equation}
  C^{\prime}=2\pi\epsilon_0\epsilon_r \left(ln(4a/d)\right)^{-1}
\end{equation} 
Taking \mbox{$a=400$\,nm} (thickness of the \mbox{SiO$_2$}),
$\epsilon_r=4.7$, a nanotube diameter of \mbox{$d=20$\,nm}
with electric contacts spaced
\mbox{$L = 200$\,nm} apart
(from edge-to-edge),
the single-electron charging
energy amounts to
\mbox{$E_c=5$\,meV}. Therefore, 
Coulomb blockade is expected to be relevant 
already for
\mbox{$T \alt 60$\,K}. As one can see from Fig.~2 this
indeed agrees with the onset of the resistance
increase. It is therefore tempting to associate
the resistance increase with charging effects
(electron-electron interaction).
True Coulomb blockade (CB), which would cause
an exponential increase of $R$, is however not observed.
We have also not found a significant gate effect
characteristic for CB. We believe
that this is due to the relatively low-ohmic contacts.
In the opposite limit of a wire which is rather strongly
coupled to the environment, CB theory
predicts deviations from the classical conductance 
with moderate temperature and voltage dependences
(power-laws)\cite{CBstrongcoupling}.
In this limit it is not possible
to separate the part that is charged from the environment.
A unified treatment is needed. 
In case of a $1d$-wire such a model is provided
by the Luttinger liquid 
(LL) theory\cite{NTLL1,NTLL2,NTLL3}.
Recently, zero-bias anomalies have been 
observed in SWNTs with
power-law dependences in agreement with
LL theory for nanotubes\cite{Bockrath}.
This experiment provides clear evidence that
long-range Coulomb interaction 
can dominate the low energy excitations
of NTs. Because
we observe similar anomalies (see Sec.~2.5)
Coulomb interaction must be considered in
order to understand $R(T)$ of MWNTs, too.

Summarizing this discussion, we have shown
that the temperature 
dependence of the
resistance
cannot be related to the specific
DOS of graphene.
It is most likely caused
by electron-electron interaction and 
(as will be shown) by
interference contribution that
are dominant at low temperature. 
This conclusion is supported by 
magnetoresistance measurements
discussed below in Sect.~2.4.

\vspace{5mm}
\noindent
{\large\em 2.3~Aharonov-Bohm oscillations in parallel field}\\

The dependence of the electrical resistance of MWNTs
in magnetic field has been studied both for parallel
and perpendicular field. The parallel field case
has recently been published\cite{BachtoldAB}.
For completeness we
summarize the main results in this section.
Figure~3 shows a typical magnetoresistance (MR) 
measurement.
On applying a field $B$
the resistance decreases. This decrease is
associated with the phenomenon of 
weak localization (WL)\cite{WLgeneral}.
WL originates from the quantum-mechanical
treatment of backscattering which contains
interference terms adding up constructively 
in zero field. Backscattering is thereby enhanced
leading to a resistance larger than the 
classical Drude resistance.
Because the interference terms cancel in magnetic
field of sufficient strength,
WL results in a negative MR.
However, for the
specific geoemetry of a cylinder (or ring), 
the WL contribution
is periodic in the magnetic flux through the cylinder
with period $h/2e$\cite{PeriodCylinder}.
Indeed, in Fig.~3 the resistance has
a second maximum at \mbox{$B=8.2$\,T}. From this field
a diameter of \mbox{$d=18$\,nm} is obtained for this 
MWNT. As was demonstrated by 
Bachtold~{\it et al.}\cite{BachtoldAB},
the MR agrees with the Altshuler, Aronov and 
Spivak (AAS) theory\cite{AAS} only, if the current is assumed
to flow through one or at most two metallic
cylinders with a diameter corresponding to the 
measured outer diameter of the NT. 
It is therefore most likely that only one cylinder
actually participates in transport. This may
not be too suprising if we consider 
the strong anisotropy
in conductivity for graphitized compounds
and the fact that
the electrodes are in direct 
contact with the outermost NT only.
The conclusion that only {\em one} graphene cylinder
carries the current can only 
unambigously be drawn from the analysis of the
low-temperature data \mbox{($T \alt 20$\,K)}.
We emphasize that it is not possible to relate
the resistance maxima at \mbox{$\pm 8.2$\,T} to
a magnetic flux of $h/e$, because a tube diameter
would then result 
which is larger than the actually measured
outer diameter. The observation
of a pronounced $h/2e$ resistance peak
proves that backscattering is present
in our MWNTs.
The NTs are therefore {\em not} ballistic.
\begin{figure}[htb]
  \epsfxsize=75mm
  \centerline{\epsfbox{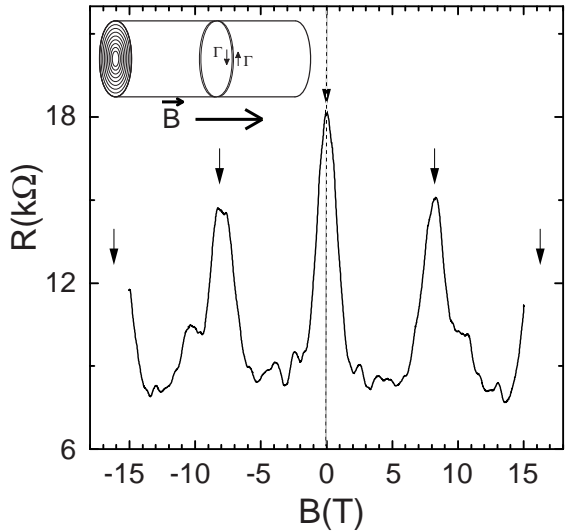}}
  \vspace{9pt}
  \caption{
  Electrical resistance $R$ as a function of magnetic field
  $B$ of a MWNT aligned parallel to $B$. Arrows denote the
  resistance maxima corresponding to multiples of
  $h/2e$ in magnetic flux through the nanotube
  taking the outer diameter.
  The contact separation is \mbox{$350$\,nm}.}
\end{figure}  

In parallel magnetic field resistance maxima
should occur periodically. Up to now, only the
first period could be observed, which is little
evidence for a periodic phenomena. 
However, in Fig.~3  the onset of
the second resistance peak is clearly seen. 
The resistance  
increases again at \mbox{$14-15$\,T}. The maximum
is expected at \mbox{$B=16.4$\,T} (see arrows), 
which is unfortunately beyond 
the field range of our magnet.

We can use Fig.~3 to estimate
the phase-coherence length. 
The zero-field resistance
peak has a full width at half maximum of
\mbox{$2\Delta B\approx 2$\,T}. 
$\Delta B$ roughly
corresponds to a flux quantum $h/e$ within an area
bounded by the wire diameter and the phase-coherence
length $l_{\phi}$.
From this condition \mbox{$l_{\phi}\approx 200$\,nm}
is obtained. As a test for consistency the WL correction
to the conductance $\delta G$ is compared with
the measurement.
For $1d$-WL $\delta G$ is of 
order $(2e^2/h)l_{\phi}/L$.
Taking \mbox{$L=350$\,nm} and 
\mbox{$l_{\phi}=200$\,nm}
we obtain \mbox{$\delta G = 4.4\cdot 10^{-5}$\,S}, which
is in very good agreement with the measured
conductance change of \mbox{$\delta G = 4.6\cdot 10^{-5}$\,S}.
From MR measurements in prependicular field
similar coherence lengths are extracted. This will
be discussed in the next section.

\vspace{5mm}
\noindent
{\large\em 2.4~Interference effects in perpendicular field}\\

Possible reasons for the measured resistance
increase at low temperature (see Fig.~2)
are interference corrections 
(WL) and electron-electron interaction effects.
In order to quantify these contributions 
an extensive investigation of the magnetoresistance (MR)
in perpendicular magnetic field was conducted.
We note from the start that the conventional
WL theory for diffusive transport is used
in the analysis of our experiments\cite{WLgeneral}.

\begin{figure}[htb]
  \epsfxsize=75mm
  \centerline{\epsfbox{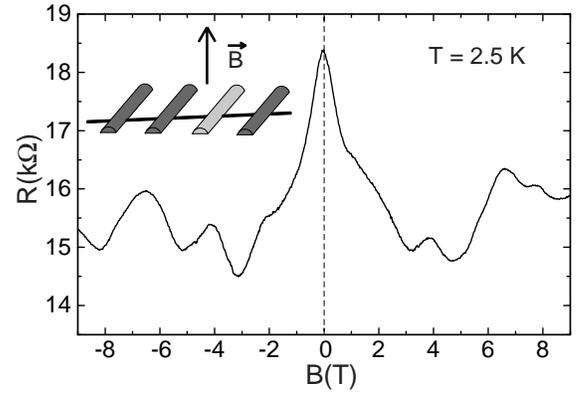}}
  \vspace{9pt}
  \caption{
  Four-terminal magnetoresistance 
  of a MWNT in perpendicular field measured
  at \mbox{$2.5$\,K}.
  The contact separation is \mbox{$350$\,nm}. The
  observed asymmetry is related to one of the
  voltage probes which was high-ohmic.}
\end{figure}  

We begin with a measurement that proves the 
presence of interference contributions. 
Figure~4 shows the four-terminal
resistance of a MWNT measured
at \mbox{$T=2.5$\,K}. Three
distinct features can be seen:
in the first place, the
resistance is largest at $B=0$. 
Secondly, there
are aperiodic resistance 
fluctuations away
from zero magnetic field, 
which are reproducible. And
thirdly, the MR is asymmetric in field, i.e.
\mbox{$R(-B) \not = R(B)$}.
All there features point to the presence
of quantum interference.
The resistance maximum at $B=0$ is caused by
weak localization as already mentioned in the 
previous section. The aperiodic fluctuations
can be assigned to direct interference contributions
and are
known as universal-conductance 
fluctuations (UCF)\cite{UCF}.
These fluctuations depend on the specific scattering
potential. They disappear if one would average
over an ensemble of otherwise similar NTs.
Particularly interesting is the third observation.
By virtue of the reciprocity theorem\cite{Buettiker}
a two-terminal
resistance must be symmetric in magnetic field.
But a four-terminal resistance, as the one measured
in Fig.~4, may be asymmetric. 
The asymmetry
arises if the voltage measured over the
two inner contacts contains 
non-local contributions
from parts of the wire that do not reside
in between the two contacts
over which the voltage is measured. 
For such non-local
contributions to occur
(a hallmark of mesoscopic physics\cite{Nonlocal})
two conditions need to be met:
1), the phase-coherence length should be 
large enough, and 2),
the voltage probes should not behave
ideally in the sense that all electrons
incident from the NT to the contacts are
absorbed with probability one.
Indeed, the contact resistance of one
voltage probe was relatively high-ohmic
for this sample. Its resistance was
\mbox{$\approx 150$\,k$\Omega$},
whereas the other contacts had 
\mbox{$\alt 10$\,k$\Omega$}.
When the temperature was increased
above \mbox{$20$\,K} (not shown), the
asymmetry as well as the aperiodic oscillations
disappeared, whereas the resistance maximum at
zero field remained present, although with
lower magnitude. Exactly this dependence is
expected once \mbox{$l_{\phi}\ll L$}.

The MR in Fig.~4 provides a convincing 
demonstration
that quantum-mechanical interference terms
strongly contribute to the electrical resistance
of carbon nanotubes at low temperature.
Let us emphasize that non-local contributions
are prohibited if all contacts are low ohmic (ideal).
This is because electrons that arrive at the
contacts are scattered with 
high probability into these where they 
are randomized.
Hence, interference effects are terminated 
at ideal contacts.
All MR measurements (also
four-terminal ones)
should therefore be symmetric in magnetic field.
Any asymmetry is a signature of a `bad' contact.
Indeed, if all contacts are low-ohmic
the resistance is found to be symmetric
in $B$ whether a $2$-terminal or $4$-terminal
measurement is conducted.

\begin{figure}[htb]
  \epsfxsize=75mm
  \centerline{\epsfbox{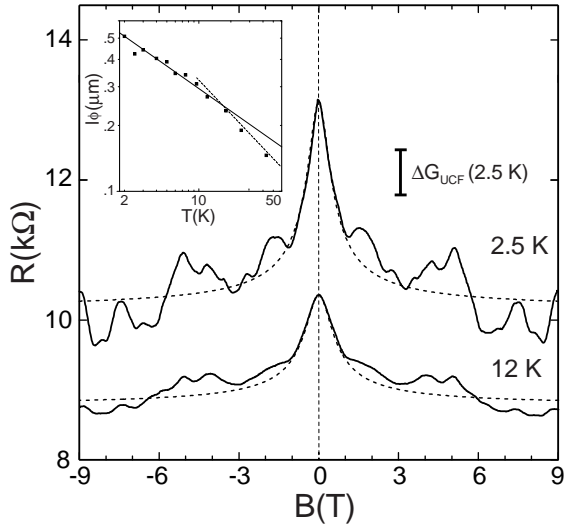}}
  \vspace{9pt}
  \caption{
  Four-terminal magnetoresistance of a MWNT in perpendicular
  field for two temperatures. The voltage probes are
  separated by \mbox{$1.9$\,$\mu$m}. Dashed curves 
  show fits using one-dimensional weak-localization theory.
  Inset: deduced phase-coherence length $l_{\phi}$
  as a function of temperature $T$.}
\end{figure}  
An example is provided by Fig.~5 which shows
the ($4$-terminal) MR of a 
MWNT for two temperatures
($2.5$ and \mbox{$12$\,K}). 
We mention that the two curves
are not displayed vertically for clarity.
This sample
differs from the previous examples in the
contact geometry.
The two inner voltage contacts are now 
separated by \mbox{$L=1.9$\,$\mu$m},
instead of \mbox{$L=0.35$\,$\mu$m} before.
The MR of this NT has been measured for
several temperatures and carefully analyzed
in order to extract the phase-coherence
length\cite{Bachtoldinprep}.
On applying a magnetic field the resistance
again decreases in agreement with WL. 
Aperiodic fluctuations are superimposed
and are assigned to UCF. The MR is compared
to $1d$-WL theory, i.e.
$l_{\phi} > \pi d$. 
We note, that our measurements
cannot be fitted by $2d$ or even
$3d$ WL, which was find to describe
the MR of NTs in 
previous work\cite{Song94,Langer,BaxendaleFujiwara,Kim98}. 
The phase-coherence length
$l_{\phi}$ is therefore larger than the diameter
in our case (\mbox{$d\approx 23$\,nm} for the 
NT of Fig.~5).

Phase-randomization is caused by all kinds
of inelastic scattering. At low temperature, however,
electron-phonon scattering can be neglected
and $l_{\phi}$ is determined by electron-electron
interaction. 
Assuming diffusive transport,
the phase-relaxation time $\tau_{\phi}$
is related to $l_{\phi}$ by 
\mbox{$l_{\phi}=(D\tau_{\phi})^{1/2}$} where
$D$ is the diffusion coefficient.
It has been pointed out that
$\tau_{\phi}$ 
has to be distinguished from the energy-relaxation
time $\tau_{ee}$ due to 
electron-electron scattering\cite{Diffintau}.
The latter is determined by scattering processes
that transfer energies of order $kT$, whereas
the phase of the wavefunction can already be 
randomized by quasi-elastic scattering events
with energy transfers $\ll kT$.
For this reason \mbox{$\tau_{\phi} < \tau_{ee}$}
in general. The temperature range over which
the last equation holds is particularly large
in $1d$. At the lowest temperatures dephasing
is always determined by quasi-elastic scattering. 
A cross-over
occurs at \mbox{$kT \sim \hbar/\tau_{ee}(T)$}.
Above this temperature the difference between
$\tau_{\phi}$ and $\tau_{ee}$ disappears.
To compare the measurements with
predictions, we use the $1d$-WL theory
which adequately takes dephasing by quasi-elastic
scattering into account\cite{Diffintau,adequateWL}. The correction
to the conductance $\delta G$ for
a wire of width $w$ is given by:
\begin{equation}
  \delta G = -0.62\frac{e^2}{\hbar L}\left(\frac{1}{l_{\phi}^2} +
  \frac{w^2}{3 l_m^4} \right)^{-1/2}
\end{equation}
Here, $l_m(B)$ is the magnetic length
given by \mbox{$l_m^2=\hbar/eB$}.
There are two important points to make: 1),
our wire is not planar but actually a cylinder,
so that $w=d$ cannot be assumed. 2),
the discussion in Sect.~2.2 has shown that
the mean-free path for elastic scattering
may be of order of the circumference of the NT.
In this case flux-cancellation due to intersecting
closed electron trajectories have to be
considered\cite{fluxcancellation}.
For these two reasons we treat the
width $w$ as an additional parameter
(in addition to $l_\phi$)
and mention that $w$ turns out to be 
$\approx d/2$.
Best fits to theory are shown in Fig.~5 by dashed lines.
A very good agreement is found. As a cross-check,
UCF amplitudes deduced using $l_{\phi}$
agree with the observation. An example
is given by the vertical bar
in Fig.~5 which corresponds to the 
expected UCF amplitude 
at \mbox{$2.5$\,K}.

There are two other subtleties 
we would like to mention here. The 
equation for $1d$-WL predicts a divergence
for $\delta G$ in the
zero-temperature limit for which 
$l_{\phi}\rightarrow \infty$ is expected. 
However, this would
only be true for an infinite long wire. 
The above equation is actually only valid 
if the length of the wire 
$\gg l_{\phi}$, because the localization
correction saturates in the limit of
$0d$\cite{WL0d}.
The second subtlety is
due to the electric contacts
which, if low-ohmic, limit the effective length
of the wire due to dephasing. Once $l_{\phi} \sim L$ 
substantial dephasing will be caused
by the contacts\cite{Diffintau,ContactDephasing}. 
Both cases predict a saturation of
the resistance increase due to weak
localization, as observed in $R(T)$.

Dephasing by the contacts can
be incorporated in to the theory
by taking the additional electromagnetic 
fluctuations of the external 
circuit into account.
The equation for $\delta G$ is thereby
modified. In order to extract 
the `intrinsic' $l_{\phi}$ 
from the measurements
we have used such a modified 
$1d$-WL equation\cite{Bachtoldinprep}.

The temperature dependent coherence length
is shown in the inset of Fig.~5.
$l_{\phi}$ is of order of a few \mbox{$100$\,nm}
in agreement with the value obtained in
the previous section using a qualitative
argument. Below \mbox{$10$\,K}, 
\mbox{$l_{\phi}$} scales with temperature
according to \mbox{$l_{\phi}\propto T^{-1/3}$}
as expected from the theory which predicts
\begin{equation}
  l_{\phi}= \left(\frac{D G_D L \hbar^2}{2e^2kT}\right)^{1/3}
\end{equation}
for dephasing by quasi-elastic 
electron-electron  
scattering\cite{WLgeneral,NyquistDephasing}.
Here, $G_D$ is the Drude conductance due to elastic scattering
only. This equation allows us to extract the
diffusion coefficient $D$ from the magnitude of
\mbox{$l_{\phi}$}.
We obtain \mbox{$D=450\dots 900$\,cm$^2$/s}.
The large range is mainly due to errors in
determining the Drude conductance. 
With a Fermi velocity of 
\mbox{$v_F=10^6$\,m/s} the elastic-scattering
length is found to be
\mbox{$l_e=90\dots 180$\,nm}.
$l_e$ is indeed very large and the values
are consistent with the estimate 
in Sect.~2.2 based on the
absolute resistance value.
Because $l_e \agt \pi d$, 
transport in our
MWNTs should be classified as
{\em quasi-ballistic}\cite{quasiballistic}
rather than diffusive.
$l_{\phi}(T)$ only follows the $T^{-1/3}$
dependence up to \mbox{$20$\,K}. Above,
$l_{\phi}$ decays faster, approximately as
\mbox{$\propto T^{-1/2}$}. Two possibilities
can account for this: 1), electron-phonon
interaction may set in, or 2), the transition
to $\tau_{\phi}=\tau_{ee}$ may occur.
Both scenario are compatible with the observed
temperature dependence within this limited
temperature interval. Because there is no
clear signature for electron-phonon scattering
in the temperature dependence of the resistance
for \mbox{$T \leq 300$\,K}, we prefer the latter
possibility. This is supported by the
temperature \mbox{($20$\,K)} for which
the transition occurs.
Taking \mbox{$D=700$\,cm$^2$/s} and
\mbox{$l_{\phi}(20$\,K$)=230$\,nm} gives
a dephasing time of \mbox{$\tau_{\phi}=0.76$\,ps}
corresponding to an energy uncertainty
of \mbox{$\approx 10$\,K} in rough agreement
with \mbox{$20$\,K}. Hence, for
\mbox{$T > 10-20$\,K} dephasing by quasi-elastic
electron-electron scattering is no longer
dominant because \mbox{$\hbar/\tau_{ee} > kT$}.
There are not 
enough data points available to
extract the temperature
dependence of $\tau_{ee}$. The 
dependence is also expected to change once
$l_e$ is no longer the shortest scattering
length (clean limit). 
For 
the sake of the 
following consideration let us
extrapolate $l_{\phi}$ to room temperature (RT)
using the observed $T^{-1/2}$ dependence.
We then obtain 
\mbox{$l_{e-e}=60$\,nm} at RT as an upper bound.
It is therefore justified to say 
that MWNTs are not
(quantum) ballistic at room temperature 
on a length scale of \mbox{$1$\,$\mu$m}. 
There is substantial 
electron-electron scattering
on a much shorter length scale. We believe that
this conclusion is valid not only for our MWNTs
but for all tubes of high-quality for which
scattering by static defects is not the dominant
scattering mechanism at RT.
Although $l_{ee}$ is $\ll L$ at RT, its effect  
on the electric resistance is not obvious.
If for example Umklapp-processes
are suppressed 
even at RT, electron-electron scattering
does not change the resistance because
the total momentum is conserved in the course
of scattering. Hence, it may still be possible
that the resistance is close to the expected
quantized value.

\begin{figure}[htb]
  \epsfxsize=75mm
  \centerline{\epsfbox{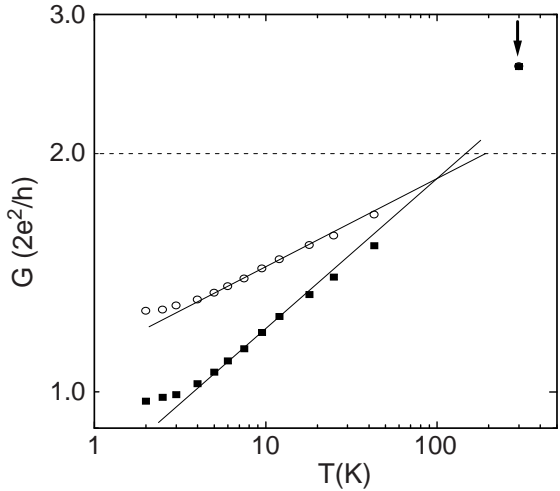}}
  \vspace{9pt}
  \caption{
  Conductance $G$ at zero magnetic field
  as a function of temperature $T$ for the
  measurement shown in Fig.~5. 
  Filled squares correspond to
  $G$ at $B=0$ and open circles to $G-\delta G_{WL}$
  with $\delta G_{WL}$ the contribution
  to the conductance from weak localization.
  The dashed horizontal line is the conductance
  expected for an ideal metallic carbon nanotube.
  The arrow points to the measured room temperature value.}
\end{figure} 
Weak localization results in an increase of
the electrical resistance at low temperatures.
This is not the only contribution to the 
resistance increase, as can be seen from
Fig.~5. For large magnetic field
$\delta G_{WL}\rightarrow 0$ but the resistance
is still seen to be strongly temperature
dependent. This temperature-dependent
background resistance is usually
associated with electron-electron interaction.
While WL primarily enters as a correction to
the diffusion coefficient,
the interaction suppresses the
single-particle DOS.
Furthermore,
interaction effects 
are enhanced 
by disorder.
For a diffusive wire for which the
coherence length (the thermal length)
is larger than the width
but smaller than the length, theory
predicts for the conductance
correction
\mbox{$\delta G_{ee}\propto T^{-1/2}$}
\cite{eeGcorrection}.
Knowing $\delta G_{WL}(T)$ one can plot
$G(T) - \delta G_{WL}(T)$ as a function of
$\sqrt{T}$. Contrary to our expectation, 
the predicted temperature dependence is
not observed. For this reason,
$G(T)$ has been plotted
in a log-log representation in Fig.~6. 
Filled symbols
correspond to the measured conductance for
zero field, whereas the open symbols represent
the conductance after subtracting $\delta G_{WL}$.
This plot is instructive. At low temperatures
\mbox{($T \alt 4 $\,K)} $G$ saturates.
In the range \mbox{$4 \leq T \leq 40$\,K} the
dependence follows approximatly a power law
\mbox{$G\propto T^p$} with 
a small exponent \mbox{$p\approx 0.1-0.2$}.
Unfortunately, no data points were measured
between \mbox{$50$\,K} and RT.
But the measured data points
show a slight negative curvature
around \mbox{$10-40$\,K} suggesting
a possible saturation at high temperatures,
which would come close to the expected
conductance of a perfect NT, i.e. $4e^2/h$.
This is -- at least qualitatively -- the $G(T)$
dependence expected for a Luttinger
liquid with some degree of backscattering\cite{Egger}.
The exponent $p=0.1-0.2$ is however relatively 
low and may indicate that the 
strong backscattering
limit is not reached.
Note, that for this sample 
the absolute value of $G$  
is surprisingly close to the
quantized conductance $2G_0=4e^2/h$ 
expected for a perfect
single-wall nanotube. That the actually measured
room temperature conductance is larger 
than $2G_0$ is not
too surprising for the following two reasons:
1), at RT it is possible that additional graphene
cylinders contribute to the measured conductance
and not only the outermost shell, as inferred from
Aharanov-Bohm measurements at low temperatures.
2), even if we stick to just one single tube higher
subbands have to be considered.
The contribution to the conductance of a 
$1d$-subband with a threshold energy $E_1$
is readily estimated to be
\mbox{$G_0 exp(-E_1/kT)$}. This has to be
multiplied with the number of available subbands,
which is four for the first higher (lower) 
subbands\cite{Datta}.
Hence, the total contribution
to the conductance is 
\mbox{$8 G_0 exp(-E_1/kT)$}, where $E_1$ is of
order \mbox{$50-60$\,meV}.
From this expression the
conductance of a single \mbox{$20$\,nm}
diameter perfect nanotube is estimated to be rather 
$3\,G_0$ instead
of $2G_0$ at RT.

\vspace{5mm}
\noindent
{\large\em 2.5~Tunneling spectroscopy}\\

\begin{figure}[htb]
  \epsfxsize=75mm
  \centerline{\epsfbox{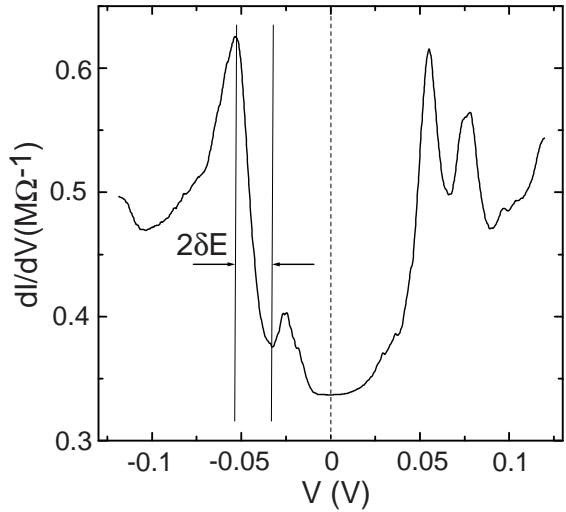}}
  \vspace{9pt}
  \caption{Differential (tunneling)
  conductance $dI/dV$ measured
  on a single MWNT using a high-ohmic contact
  \mbox{($300$\,k$\Omega$)}
  at \mbox{$T=4.2$\,K}. This spectra confirms
  to the DOS expected for a metallic
  nanotube in which the wave vector is quantized 
  around the tube circumference leading 
  to $1d$-subbands. $\delta E$ denotes the
  sharpness of the the observed van-Hove singularities.
  Positive (negative) voltages correspond to empty
  (occupied)
  nanotube states.}
\end{figure} 
Figure~7 shows a differential current-voltage
characteristic \mbox{($dI/dV$)} measured
on a single MWNT using an inner contact
which by change was high ohmic. This particular
tunneling contact had \mbox{$300$\,k$\Omega$}
whereas the other contacts had resistances
\mbox{$\ll 10$\,k$\Omega$}. 
The measured
spectrum agrees surprisingly well with
predicted spectra based on simple tight-binding
calculations for a metallic NT\cite{Datta,DOStightbinding}.
Firstly, there
is a substantial DOS at the 
Fermi energy, i.e. at $V=0$, so that the
NT is metallic. Secondly, the almost symmetric
peak structure, appearing as a pseudogap
is caused by the additional $1d$-subbands
in the valence ($V < 0$) and conduction band ($V > 0$)
with threshold energies of order \mbox{$\approx 50$\,meV}.
At the onset of the subbands van-Hove
singularities are expected. The spectrum
in Fig.~7 agrees remarkably well with
the scanning-tunneling measurements of
Wild\"oer~{\it et al.} for SWNTs\cite{Wildoer}.
But because
of the difference in tube diameter the energy scales
are different. While the distance beween the two
first-order subbands $\Delta E_{sb}$ is found to 
\mbox{$\Delta E_{sb}=1.8$\,eV} for a SWNT with diameter
\mbox{$d=1.3$\,nm}, we found
a smaller value of \mbox{$0.12$\,eV}.
This is reasonable, since $\Delta E_{sb}$ should be
proportional to 
$d^{-1}$\cite{AijkiAndo,DOStightbinding}.
According to this 
relation, a diameter
of \mbox{$19$\,nm} is predicted for
our MWNT in good agreement
with the measured outer diameter of
\mbox{$d=17$\,nm}. 
This is another independent proof of the
conjecture that the electric contacts 
probe the outermost tube of a MWNT. 
Moreover, the observed 
occurence of $1d$-subbands
in $dI/dV$ shows that the motion 
of electrons in our NTs cannot be
regarded as  
$2d$-diffusive but must be ballistic
on the scale of the tube diameter. 

The van~Hove singularities are broadened by
thermal smearing as well as scattering.
The limiting factor is scattering because
\mbox{$kT \ll \delta E\approx 10$\,meV} for
the measuring temperature which was 
\mbox{$T=4.2$\,K}. 
An alternative estimate
for the scattering mean-free length
can now be given.
The $1d$-subband can only develop
as a pure eigenstate if $l_e \gg \pi d$.
If we relate the scattering time to $\delta E$
we obtain \mbox{$l_e \alt 150$\,nm}.
This relatively large elastic-scattering length 
supports the previous 
quantitative WL analysis which
predicted similar lengths, i.e.
\mbox{$l_e=90\dots 180$\,nm}.
These two consistent results convincincly establish
that (our) MWNTs are quasi-ballistic conductors.

The observed spectrum in Fig.~7 
nicely demonstrates that the 
peculiar bandstructure
effects of NTs are also found for MWNTs.
We have to emphasize, however, that a spectrum
with sharp van-Hove singularities 
in close agreement with tight-binding 
calculations has only been
observed on one sample until now, although
several MWNTs have been studied.
The prevailing spectra display a pronounced
zero-bias anomaly on a smaller energy scale
of \mbox{$1-10$\,meV}. For larger energies,
they resemble Fig.~7 in the sense that a peak-structure
develops in $dI/dV$ on the 
scale of the subband separation 
\mbox{($0.1$\,eV)} 
which may be associated
with (broadened) van-Hove singularities.

\begin{figure}[htb]
  \epsfxsize=75mm
  \centerline{\epsfbox{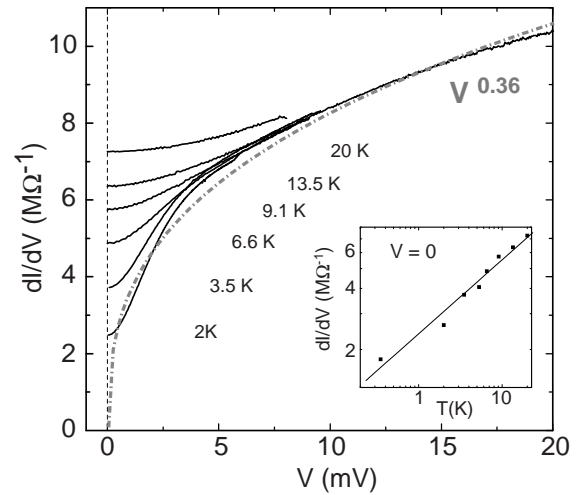}}
  \vspace{9pt}
  \caption{ Differential tunneling 
  conductance $dI/dV$ measured
  on a single MWNT at different temperatures $T$
  displaying a pronounced zero-bias anomaly
  on a relatively small voltage scale as compared
  to Fig.~7.
  Inset: log-log
  representation of $dI/dV$ vs. $T$ for
  $V=0$. The dashed-dotted curve displays the power-law
  $dI/dV \propto V^{\alpha}$ with $\alpha = 0.36$
  deduced from the inset.}
\end{figure} 
A typical zero-bias anomaly (ZBA) is shown in
Fig.~8 for six temperatures ranging from
\mbox{$2-20$\,K}. We note, that the full 
$dI/dV$ displays a slight asymmetry in voltage
and the data in Fig.~8 correspond to the symmetric part.
A suppression of the tunneling DOS
is expected for a strongly correlated 
electron gas\cite{LLreview}.
Similar anomalies have recently been observed
by Bockrath~{\it et al.} for 
SWNTs\cite{Bockrath}. Their 
measurement and analysis provide the first
demonstration for Luttinger liquid (LL)
behavior in carbon NTs due to long-range
Coulomb interactions. The LL liquid theory
describes the interaction 
with a single parameter
$g$\cite{LLreview}. 
The non-interacting
Fermi-liquid case
corresponds to $g=1$ and $0<g<1$
is valid for a LL with 
repulsive interaction. The parameter $g$ is 
determined by the ratio of the single-electron
charging energy to the single-particle level
spacing and has been estimated to be 
\mbox{$g=0.2-0.3$} for SWNTs\cite{NTLL3}. 
Because of the
larger diameter of MWNTs, $g$ may even approach
a value of $\approx 0.35$ for a single shell
of diameter \mbox{$30$\,nm}. However, the exact
value is difficult to estimate because the
assumption of a single shell may be wrong 
in case of a MWNTs. Though contributing 
little to transport,
the inner shells of a MWNT may strongly
screen the long-range Coulomb
interaction leading to $g \rightarrow 1$.
It may be quite possible that the 
single-particle like
$dI/dV$ spectrum of Fig.~7 is due to 
such a screening effect.
Pressuambly, this is also the reason
why a single particle like DOS
has been measured on SWNTs in
good contact to a Au substrate\cite{Wildoer},
although LL effects should be most prominent
in unscreened SWNTs on an insulating
substrate.

LL theory predicts power laws both
for the voltage and temperature dependence,
i.e. \mbox{$dI/dV(T,V=0)\propto T^{\alpha}$} and
\mbox{$dI/dV(T=const,V) \propto V^{\alpha}$}
if \mbox{$eV \gg kT$}\cite{LLreview}.
The exponent $\alpha$ is related to $g$, explicitly
\mbox{$\alpha = (g^{-1}+g-2)/8$} for tunneling
into the bulk appropriate in our situation\cite{NTLL3}.
A power-law with $\alpha \approx 0.36$ is
deduced from \mbox{$dI/dV(T,V=0)$}; see the inset
of Fig.~8. For comparison with the observed
$dI/dV$-voltage dependence, the dashed-dotted
curve $\propto V^{0.36}$ has been
plotted. The agreement with the measurements
is not perfect, but taking into account the
above mentioned asymmetry quite satisfactory.
The asymmetry is possibly caused by 
bandstructure effect. \mbox{$\alpha = 0.36$}
corresponds to \mbox{$g=0.21$}. The same value
was obtained by Bockrath~{\it et al.}\cite{Bockrath}.
This exact agreement has
presumably no significance because
we use single MWNTs whereas they have used SWNT ropes.
On the other hand, the agreement may indicate
that the same physics is responsible for the ZBA.
One has to keep in mind that
the single-particle DOS can
be suppressed for other reasons 
like the presence of two-level and
multi-level systems.
However, we have not observed
temporal fluctuations characteristic
for such multi-level systems. 
Furthermore, the observed ZBA is
not of magnetic origin, because
we observe no significant change in a 
perpendicular magnetic field up to 
\mbox{$12$\,T}.

\vspace{5mm}
\noindent
{\large\bf 3~Critical discussion}\\

Because resistivities were reported in previous
studies using films of NTs we first like to make
a comparison with these results.
A typical value determined for the 
resistivity $\rho$
is \mbox{$10^{-3}$\,$\Omega$cm}\cite{Song94,Kim98,NTfilms}.
Using the resistance that we have been measuring and
the knowledge that the electric current flows
through the outermost cylinder of the MWNT,
we estimate \mbox{$\rho \approx 10^{-6}$\,$\Omega$cm}.
This large difference indicates that the volume fraction
of conducting NTs in thin-film samples is low.
For this reason, it is not possible to compare
the thin-film results with measurements on single
NTs, because extrinsic effects (like intertube hopping)
most likely dominate the resistance of thin-film samples.

For all measured MWNTs we find a
resistance of order \mbox{$10$\,k$\Omega$}
at low temperature and a similar
temperature dependence.
The resistance increases by a factor $\approx 2-3$
if the temperature is lowered from RT to
a few K. In addition, interference effects
show up at low temperatures indicative for
the presence of backscattering. In interpreting
these results we have extensively used
conventional weak-localization (WL) theory.
This theory relies on diffusive transport
in $2d$, i.e. $l_e$ is implicitly assumed to be 
much smaller than the cirumference of the NT.
It may appear as a contradiction that using
this theory $l_e$ turns out to be larger
than \mbox{$\pi d$} so that a central assumption 
for the validity of the theory is not met.
In one example, tunneling spectroscopy has revealed 
a spectral density 
that agrees with 
bandstructure calculations
for which $l_e > \pi d$ is required. 
Furthermore, comparing the typically 
measured resistance value with a Drude expression
comparable values for $l_e$ are found.
These additional results give us confidence in the
magnitude of $l_e$.
MWNTs are therefore rather ballistic than that
they are diffusive. Strictly speaking, they are
{\em quasi-ballistic}. 
However, we do not think that
this conclusion contradicts
the observed interference corrections, 
because interference effects
have also been seen in split-gate 
quantum-point contacts (QPC)
which are the prototype 
ballistic device\cite{interferenceQPC}.
In order to obtain the quantized conductance,
theory assumes that the $1d$-subbands in the
constriction of the QPC are adiabatically
coupled to states of the reservoirs on either side
leading to zero backscattering\cite{adiabaticQPC}.
In real devices
perfect adiabaticity is never possible. Still,
nice quantized conductance plateaus can be observed
and the QPC appears to be ideal. However,
if the temperature is lowered enough resonances
appear in particular in the vicinity of
subband thresholds. 
The conductance vs. gate dependence appears
more and more `messy' the lower the temperature.
This is because of the increase in 
phase-coherence length which can lead to
pronounced interference contributions in the 
resistance due to residual random
potential fluctuations. Loosly speaking one may state that
the QPC appears much more ideal at higher temperatures.
The same may be true for NTs.
Let us denote the mean fluctuation of the 
(self consistent) potential by $\delta E_d 
\sim \sqrt{\overline{\delta^2 V}}$.
The appropriate energy-scale with which
to compare \mbox{$\delta E_d$} is the 
thermal energy $kT$\cite{Egger}.
If $kT \gg \delta E_d$ scattering is weak and the
system appears ideal. The conductance would be 
quantized if backscattering can be neglected all together.
In contrast, if the temperature is lowered such that
\mbox{$kT < \delta E_d$} the strong scattering regime
is entered and resonances and the like are expected.
Qualitatively, this is what we observe.

We note, that even at low temperature
backscattering cannot be very
strong. Based on the theoretically
expected ideal conductance, the total transmission 
probability at low temperature is still
reasonably large 
amounting to \mbox{$\approx 0.5$}. 
This has to be
the case, because we have never observed
a transition to strong localization. 
The source of the
remaining scattering is not clear. The electric 
contacts are certainly not ideal, even if apparently
low ohmic. Scattering at the contacts alone is
not sufficient to cause the aperiodic resistance
fluctuations (UCF) seen at low temperatures. Therefore,
there is some weak backscattering inherent to the
MWNTs. This is also supported by comparing the
resistances measured on one specific NT 
for different contact spacings.
The resistance is larger for larger contact
separations. 
The relatively large transmission probability and
elastic-scattering length together with a disorder potential
smaller than the subband separation (as suggested by Fig.~7)
point to the high quality of the MWNTs that we are using.
 
Although the nanotubes are nearly ideal, the
interference corrections (WL and AAS) can nicely
be fitted by the traditional theories despite
the fact that the theories are rigorously
speaking no longer valid. The large range
of (approximate) validity of the theory, 
which to us came 
as a surprise, is possibly just related to the
the well accepted universality of interference
corrections both in magnitude and 
magnetic-field dependence. 
For the case of magnetoresistance
in perpendicular field, we can imagine that
the resistance corrections resemble weak localization
even for few transport modes only.
In contrast, for the parallel field case there
is a serious problem. On the one hand, the
AAS correction is based on time-reversed trajectories
due to scattering along the circumference of the
tube, which requires $l_e \alt \pi d$. This is
just the opposite of what we have infered before.
On the other hand, if we assume that the
electronic structure has to be treated
one-dimensionally, the bandstructure 
is predicted to be periodically
modulated with period $h/e$
if the interior of a carbon NT is 
threaded by a Aharonov-Bohm flux\cite{Datta,ABband}.
A metallic nanotube
would be turned into a semiconducting one with a
considerable band-gap of order \mbox{$0.1$\,eV}.
Although the magnetotransport experiments in
parallel field have been performed
for much lower temperatures, a transition to
an insulating state has never 
been observed\cite{BachtoldAB}.
This is puzzling, since
$dI/dV$ (Fig.~7) suggest that MWNTs can 
be described by the simple tight-binding 
bandstructure.
This is a problem which deserves 
more attention
in the future. 
Until now, magnetoresistance
and $dI/dV$ have been measured
on different NTs.
It would be 
highly desirable
to obtain a complete set of
data for one NT including
$dI/dV$, $R(T)$, and MR in parallel
and perpendicular field.

Finally, we like to mention another point which
need to be addressed by theory.
For quantum wires fabricated from semiconducting
heterostructures it is well known that backscattering
is reduced if a perpendicular magnetic field
of sufficient strength is applied\cite{QPC}.
In the limit of large fields, $1d$-subbands
are formed which are localized along the edges, so-called
edge-states\cite{edgestatepicture}.
Because modes at opposite edges
propagate in different directions (are chiral) 
the suppression is very effective leading to
the ideal quantization as observed in the integral
quantum Hall-effect. 
For NTs
it has been shown that the wavefunctions of NTs
are shifted to the sides in perpendicular 
magnetic field, so that 
edge states are formed\cite{AijkiAndo}.
In MWNTs this occurs
already for a relatively low magnetic field
of order \mbox{$1$\,T}, because the
magnetic length \mbox{$l_m = \sqrt{\hbar/eB}$}
at \mbox{$1$\,T} is \mbox{$l_m = 26$\,nm} and
therefore comparable to the (outer-) diameter
of our MWNTs.
We think it is important to theoretically study
how the formation of edge-states influences 
backscattering in the presence of weak disorder.
A first treatment of this problem concludes
that the MR should be positive\cite{posMRedgestates}, 
but this is
not observed.

Similar to SWNTs, MWNTs are nearly ideal
conductors with only two occupied $1d$-subbands.
Because the long-range Coulomb interaction
is expected to be strong, the Luttinger liquid (LL)
picture is presumably the adequate description.
This is supported by the observed zero-bias anomaly
as well as the temperature-dependent conductance
which cannot be accounted for by conventional
Fermi liquid theory. We think that MWNTs are LL
with some degree of (weak) disorder.
Assuming good (ideal) contacts
theory predicts \mbox{$G\rightarrow 2G_0$}\cite{Maslov,Egger}
if \mbox{$kT \gg \delta E_d$} and
\mbox{$G\rightarrow 0$} in the opposite limit,
provided the wire is infinitely long\cite{Egger}.
Since \mbox{$\delta E_d  \approx kT$} at RT,
$G \alt 2G_0$. Let us assume for
the argument that $G = 2G_0$ at RT.
This conductance corresponds to the two subbands
at the Fermi energy.
As has already been mentioned, the higher subbands
also contribute to $G$ at RT.
Actually, one would rather expect to measure
a conductance of $G \approx 3 G_0$ instead
of $2 G_0$. Let us now 
look at the average room temperature resistance
values of recently measured MWNTs with very good
contacts, i.e. $R_{2t} \approx R_{4t}$ to
within $20\,\%$.
For a contact separation of \mbox{$350$\,nm}
we obtain \mbox{$R_a=3.2 \pm 1.6$\,k$\Omega$}
whereas \mbox{$R_b=9.6\pm 2$\,k$\Omega$} for
a separation of \mbox{$1.9$\,$\mu$m}. 
Note, that there is a length
dependence of order \mbox{$4$\,k$\Omega$/$\mu$m}.
Extrapolating these two values to zero contact
separation gives \mbox{$R(0)\approx 2$\,k$\Omega$}
corresponding to $6 G_0$. This is larger
than $3G_0$ and suggests that
more than one tube contributes 
to the conductance at RT.
The most important remark 
we like to make here
is concerned with the results 
of Frank~{\it et al.}\cite{Frank}.
who claim a universal conductance of
\mbox{$G_0$} for similar MWNTs at RT.
This is definitely inconsistent with our
observation. Because we find a nanotube
conductance which is {\em larger} by a factor
$6$, our MWNTs cannot be blamed to be
more dirty or more disordered. Actually,
one may conclude the opposite.
At present, we cannot offer a solution for this
discrepancy.
 
To support ballistic transport over micrometer
distances, Frank~{\it et al.} came up with
another interesting experimental observation.
Large electric currents can be driven through MWNTs
without destroying them. Based on the electric
power and bulk heat conductivity for graphite
the NT is expected to evaporate due to the large
temperature rise. But this is not observed.
This observation may however not be taken
as a proof for ballistic transport. Rather
it shows that dissipation is 
largely absent (which is
an exciting fact by itself).
We precisely know that our NT are {\em not} ballistic
but still we are also able to pass 
large currents of similar magnitudes
\mbox{(mA)} through our MWNTs.
An interpretation of this phenomena is difficult
because it occurs in the non-linear transport regime
for applied voltages much larger than
the subband separation.

\vspace{5mm}
\noindent
{\large\bf 4~Conclusions}\\

The reported study of electric transport
of single MWNTs gives rise to results which appear to be
in contradiction. For example, the observation of
an Aharanov-Bohm effect with period $h/2e$ suggests 
diffusive transport on the scale of the circumference
of the nanotube, i.e. $l_e \alt \pi d$. On the other hand,
we have observed a $dI/dV$ spectrum which agrees with
tight-binding models assuming the
existence $1d$-subband. This just suggests
the opposite, i.e. $l_e \agt \pi d$. A large
elastic-scattering length is also derived 
when comparing
the measured resistance with a 
simple Drude-type equation.
All our results can therefore be
consistent only if $l_e$ is of the order of the
circumference; not very much larger, but 
also not very much smaller. 
Transport has therefore
to be characterized as {\em quasi-ballistic}.
What remains to be determined is the source 
of backscattering which at present is not known.
Possible sources are static potential fluctuations
due to adsorbates on the outer surface of the
MWNT or potential variations caused by 
a partial wavefunction overlap between 
states of the outermost and next inner graphene
cylinder.

There is a second `contradiction' 
inherent
to our presentation. 
On the one hand, we have
extensively used theories like weak-localization
which are based on Fermi liquid (FL) hypothesis.
On the other hand, the observed suppression
of the single-particle of states 
(the ZBA) suggests that NTs may develop
a Luttinger liquid (LL) state.
It is quite uncomfortable
to describe experiment A
with FL theory and 
experiment B with LL theory. 
This is only
justified because of a lack in 
theories which can be applied
to analyse the measured data.
If Luttinger liquid is the correct
description for NTs we need
to know how the observed quantum 
interference corrections have to be
described. Is there something
similar as weak-localization in the
LL picture? What happens in a magnetic field?

\vspace{7mm}
\noindent
{\it Acknowledgements}

The nanotubes used in this works were kindly provided
by L.~Forr\'o, J.-P.~Salvetat and J.-M.~Bonard. 
In addition to the authors, the following persons contributed
to this work:
M.~R.~Buitelaar, A.~Genkinger, and T.~Nussbaumer. 
We have been profited from enspiring recent discusions with
P.~Avouris, D.~H.~Cobden, L.~Forr\'o, W.~A.~de~Heer, and
E.~Sukhorukov.
This work has been supported by the
Swiss National Science Foundation.


\end{document}